\documentclass[twocolumn,epjc3]{svjour3}
\smartqed  
\RequirePackage{graphicx}
\graphicspath{{fig/}}
\usepackage{siunitx}
\RequirePackage{booktabs}
\RequirePackage[numbers,sort&compress]{natbib}
\RequirePackage[colorlinks,citecolor=blue,urlcolor=blue,linkcolor=blue,bookmarksdepth=3,bookmarksopenlevel=3]{hyperref}
\usepackage[british]{babel}
\usepackage{hhline}
\usepackage{multirow}
\usepackage{hyperref}
\journalname{Eur. Phys. J. C}

\begin{document}

\title{Characterization of High-Purity Germanium Detectors with Amorphous Germanium Contacts in Cryogenic Liquids}

\titlerunning{Amorphous Germanium in Cryogenic Liquids}

\author{R. Panth\thanksref{usd}
\and
J. Liu\thanksref{usd, e1}
\and
I. Abt\thanksref{mpi}
\and
X. Liu\thanksref{mpi}
\and
O. Schulz\thanksref{mpi}
\and
W.-Z. Wei\thanksref{usd}
\and
H. Mei\thanksref{usd}
\and
D.-M. Mei\thanksref{usd}
\and
G.-J. Wang\thanksref{usd}
}

\thankstext{t1}{This work was supported by NSF OISE-1743790, PHYS-1902577, OIA-1738695, DOE FG02-10ER46709, the Office of Research at the University of South Dakota and a research center supported by the State of South Dakota.}
\thankstext{e1}{Corresponding author E-mail address: \href{mailto:jing.liu@usd.edu}{jing.liu@usd.edu}}


\institute{University of South Dakota, 414 East Clark Street, Vermillion, SD 57069, USA\label{usd} \and
Max-Planck-Institut f\"ur Physik, F\"ohringer Ring 6, D-80805 M\"unchen, Germany\label{mpi}
}

\date{Received: date / Accepted: date}

\maketitle
\begin{abstract}
For the first time, planar high-purity germanium detectors with thin amorphous germanium contacts were successfully operated directly in liquid nitrogen and liquid argon in a cryostat at the Max-Planck-Institut f\"ur Physics in Munich. The detectors were fabricated at the Lawrence Berkeley National Laboratory and the University of South Dakota, using crystals grown at the University of South Dakota. They survived long-distance transportation and multiple thermal cycles in both cryogenic liquids and showed reasonable leakage currents and spectroscopic performance. Also discussed are the pros and cons of using thin amorphous semiconductor materials as an alternative contact technology in large-scale germanium experiments searching for physics beyond the Standard Model.

\end{abstract}

\section{Introduction}
\label{intro}

If the decay of heavy Majorana neutrinos~\cite{panella2002signals,almeida2000signature,das2018heavy,rodejohann2002phenomenological} in the early universe into leptons and antileptons created a slight matter and antimatter asymmetry~\cite{asaka2005numsm,asaka2005numsm1}, the observed asymmetry in our current universe can be explained with the help of Leptogenesis~\cite{fong2012leptogenesis,davidson2002lower}, which is a theory that converts the lepton-antilepton asymmetry to a baryon-antibaryon asymmetry. The existence of heavy Majorana neutrinos is predicted by the seesaw mechanism~\cite{foot1989see,mohapatra1980neutrino,cai2018lepton} to explain the tiny masses of the observed neutrinos compared to other leptons, such as electrons. Light neutrinos must also be of Majorana type in the scenario of the seesaw mechanism. In this case, neutrinos are their own antiparticles, and neutrinoless double-beta ($0\nu\beta\beta$) decay~\cite{dolinski2019neutrinoless,giuliani2010searches} becomes possible.

\begin{figure}[htbp]\centering
  \includegraphics[width=0.9\linewidth]{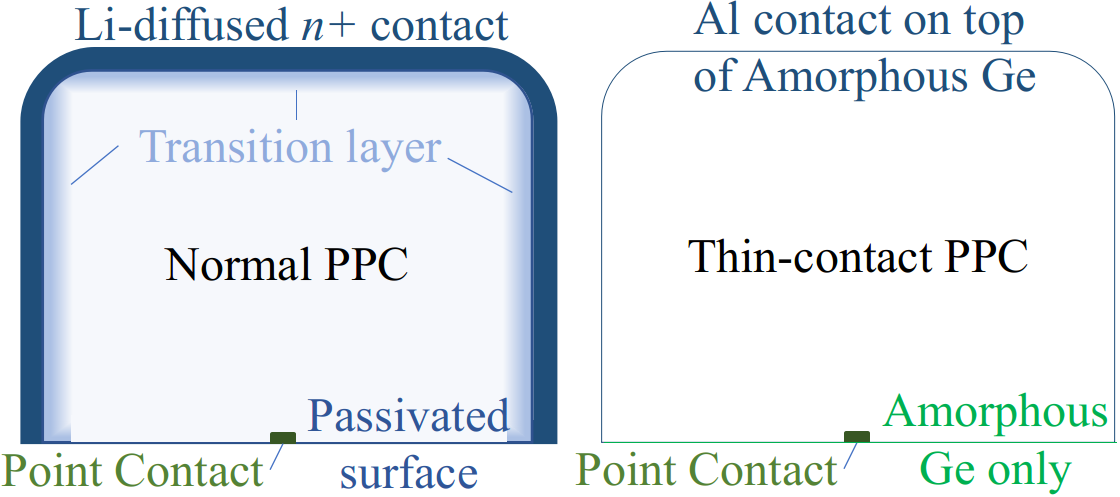}
  \caption{Comparison between a normal and a thin-contact PPC HPGe detector (not to scale).}
  \label{f:PPC}
\end{figure}

In GERDA~\cite{agostini2018improved,agostini2018upgrade, agostini19}, an experiment searching for the $0\nu\beta\beta$ decay of $^{76}$Ge, and the follow-up experiment LEGEND~\cite{abgrall2017large}, a merger of GERDA and Majorana Demonstrator~\cite{aalseth2018search, mjd19}, high-purity germanium (HPGe) detectors are operated directly in liquid argon (LAr), acting as a coolant, a passive radioactive background shielding, and an active background veto.  The detectors deployed are mostly $p$-type point-contact (PPC) HPGe detectors~\cite{giovanetti2015p,mertens2019characterization} and broad-energy germanium (BEGe) detectors~\cite{barrientos2011characterisation,agostini2011characterization} with most their surfaces being lithium-diffused contact layer as shown in Fig.~\ref{f:PPC} left~\footnote{The inverted-coaxial PPC will be used in LEGEND. It features a bore hole on the opposite side of the point-contact, which allows it to be depleted at a relatively low voltage even its overall volume is much larger than a normal PPC. However, since such a configuration does not change the discussion hereafter, it is not shown in Fig.~\ref{f:PPC} for simplicity.}. The layer is typically 1~mm thick, reducing the active volume substantially, especially when the transition region underneath the lithium-diffused layer is taken into account~\cite{de72, padraic13, aguayo13, giovanetti2015p, jiang16}. To illustrate, consider a small PPC detector with a diameter of 3~cm and a height of 3~cm, the lithium-diffused layer and the transition layer beneath it may take up to 26\% of the overall volume. The number drops to about 9\% for a detector with a diameter of 8.4~cm and a height of 10~cm, which is still a non-negligible fraction considering the price of a $^{76}$Ge-enriched PPC.

An additional consideration are background events mimicking $0\nu\beta\beta$ decays, induced by electrons from beta decays on the surface of the detector.  An example of these are  $^{42}$K (daughter of $^{42}$Ar) decays with a Q-value of 3525~keV, which can be recorded with an energy in the region of interest around 2039~keV (Q-value of $0\nu\beta\beta$ decay of $^{76}$Ge) due to partial charge collection in the outer layers of the detector.  An artificial enlargement of the lithium-diffused layer has been discussed but that would lead to a further loss in active volume.

An attractive alternative are thin contacts as shown in Fig.~\ref{f:PPC} right.  There are already commercial PPC detectors with their end surfaces made of thin contacts that are sensitive to $\alpha$, $\beta$, and low energy $X$-rays~\cite{llacer1977entrance}. Should the entire lithium-diffused contact be replaced by such a thin one, the sensitive volume of a large PPC can be enlarged by about 9\%, which is favorable for the tonne-scale LEGEND experiment.

Since the thin contact is sensitive to $\alpha$ and $\beta$ particles, such a technique has to be combined with the use of underground argon~\cite{alexander2019low}, careful selection of materials close to the detector, avoidance of surface contamination, and an active veto using LAr scintillation light.

Thin contacts can be easily segmented. Signals from a surface segment normally have worse energy resolution than those from the point-contact due to the larger capacitance of the segment. However, they can be used to precisely determine the start time of an event in a PPC, especially of an event close to the surface. Combining the time information from segments and the energy information from the point-contact, better identification of surface events may become possible.

More contacts call for more readout cables and front-end electronics, which may bring in more background. Once the number of segments becomes too large, the background induced may cancel out the benefits. Detailed Monte Carlo studies are needed to design an optimized segmentation scheme. A simple scheme for a PPC detector would be a segment for its side surface and another for the end surface opposite to the point-contact side.

A mature technique to make thin contacts is to sputter germanium or silicon on bare HPGe crystals followed by the deposition of a thin layer of aluminum to form electrodes~\cite{luke1992amorphous, luke1994140, luke2000germanium, amman2000position}. The sputtered germanium forms an amorphous germanium layer, which is about a few hundred nanometre thick. It can block the injection of both electrons and holes from contacts to the bulk of a detector, while allows charge carriers from the bulk to be collected on contacts~\cite{hansen1977amorphous}. In the surface area not covered by aluminum electrodes, it works as a passivation layer to protect the crystalline HPGe beneath.

The technique has been used to produce large planar strip HPGe detectors by Mark Amman at the Lawrence Berkeley National Laboratory (LBNL)~\cite{amman2018optimization} to detect soft $\gamma$-rays (0.2--5~MeV) in the COSI~\cite{chiu2015upcoming, kierans20172016} experiment. The properties of thin contacts have been thoroughly examined~\cite{kierans20172016} and have survived very harsh operating environments, including a crash-landing of a COSI balloon.

A dozen mini planar detectors have been fabricated at the University of South Dakota (USD) using the technique developed at LBNL and HPGe crystals produced at the USD crystal pulling facility~\cite{meng2019fabrication}. They perform well in a traditional vacuum cryostat~\cite{wei2018investigation}. A cryostat, Gerdalinchen II~\cite{gerdalinchen}, has been developed at the Max-Planck-Institut (MPI) f\"ur Physik in Munich to study segmented HPGe detectors directly submerged in cryogenic liquids, including liquid nitrogen (LN$_2$) and LAr. Reported in this paper is work carried out in summer 2019 to study the feasibility of operating HPGe detectors with thin amorphous germanium surfaces directly in LN$_2$ and LAr.

\section{Experiment}
\subsection{USD Detectors with Amorphous Germanium Surfaces}
Three mini planar HPGe detectors with amorphous germanium surfaces were used in this study. They were made from HPGe crystals grown at USD. Their dimensions and properties are summarized in Table~\ref{t:det}.

\begin{table}[htbp]
  \caption{Summary of USD detector properties.}\label{t:det}
  \begin{tabular}{lccc}\toprule
    Detector & USD-RL & $^\dagger$USD-8-4-15 & USD-R02 \\\midrule
    $^\triangleright$Impurity/cm$^3$ & $6.2\times10^9$ & $1.7\times10^{10}$ & $2.9\times10^{10}$ \\
    Thickness/cm & 1.07 & 0.70 & 0.65 \\
    Top area/cm$^2$ & $1.88\times1.79$ & $1.27\times 1.20$ & $^\ddagger 0.5\times 0.5$\\
    $^\star V_d$/V & 400 & 400 & 700 \\
    $^{\diamond}I_\text{before}$/pA & 10 & 1 & $^\ddagger$1 \\
    $^{\diamond}I_\text{LN$_2$}$/pA & 3--5 &$\leq0.2$ & $^\ddagger$1 \\
    $^{\diamond}I_\text{LAr}$/pA & 210--234 & 10 & $^\ddagger$25\\
    $^{\diamond}I_\text{after}$/pA &7  & $^\oplus$- & $^\ddagger$3 \\
    $^{\bullet}\Delta E_\text{pulser}^\text{before}$/keV & 1.93 & 1.28 & 1.67 \\
    $^{\bullet}\Delta E^\text{before}$/keV & 2.55 & 1.66 & 2.16 \\
    $^{\bullet}\Delta E_\text{pulser}^\text{LN$_2$}$/keV & 5.63 & 5.64 & $\odot$ - \\
    $^{\bullet}\Delta E^\text{LN$_2$}$/keV & 5.92 & 5.81 & $\odot$ - \\
    $^{\bullet}\Delta E_\text{pulser}^\text{LAr}$/keV & 5.44 & 4.95 & 5.42 \\
    $^{\bullet}\Delta E^\text{LAr}$/keV & 5.91 & 5.03 & 6.01 \\
    $^{\bullet}\Delta E_\text{pulser}^\text{after}$/keV & 1.10 & $^\oplus$- & 2.00 \\
    $^{\bullet}\Delta E^\text{after}$/keV & 1.74 & $^\oplus$- & 2.98 \\
\bottomrule
  \end{tabular}
  $^\triangleright$ Net impurity concentration calculated using Eq.~\ref{e:imp}.\\
  $^\dagger$ Made by Mark Amman at LBNL in 2015.\\
  $^\ddagger$ Values are for the central contact.\\
  $^\star$ $V_d$: Depletion voltage.\\
  $^\diamond$ $I$: leakage current measured at 1200~V in LN$_2$, LAr, and vacuum before/after the MPI deployment\\
  $^\bullet$ $\Delta E$: energy resolutions of the pulser and the 662 keV $\gamma$-ray peak measured at 1200~V in LN$_2$, LAr and vacuum before/after the MPI deployment.\\
  $^\oplus$ No measurement at USD after its deployment at MPI since the detector was left at MPI.\\
  $^\odot$ No measurement since the $^{137}$Cs source was temporarily unavailable.
\end{table}

Cylindrical HPGe crystal boules from Czochralski pullers operated at USD were first diced into about $2 \times 2 \times 1$~cm$^3$ cuboid with diamond wire saws and grinding blades. One of them was sent to LBNL, where the detector USD-8-4-15 was fabricated. The detailed fabrication process of this detector is described elsewhere~\cite{amman2018optimization}. At USD, each cuboid was further ground into a top hat shape, as shown in Figs.~\ref{f:setupUSD} and \ref{f:circuit}. The brims were used to handle the crystals so that their sensitive surfaces were kept untouched during fabrication and operation.  The top and bottom surfaces of the crystals were lapped using silicon carbide and aluminum oxide with 17.5 and 9.5 micron grids, respectively, to remove visible scratches from cutting. All pieces were then submerged in a mixture of HF and HNO$_3$ acids to etch away small surface defects. After rinsed in de-ionized water and dried with nitrogen gas, all surfaces were shiny and reflective.

Amorphous germanium was deposited on all surfaces in a radio-frequency sputtering machine. The sputtering was done in a 93:7 mixture of Ar and H$_2$ gas at 14~mTorr. The duration of the sputtering was carefully controlled such that the thickness of the amorphous germanium layers became about 300~nm. Aluminum contacts were then evaporated on the top and bottom surfaces using an electron-beam evaporator for the detectors USD-R02 and USD-8-4-15. For the detector USD-RL, the aluminum contacts were sputtered on. Any undesired deposition of aluminum on the side surfaces was etched away in a 1\% HF solution. The final contact structure is sketched in Fig.~\ref{f:circuit}.  The fabrication procedure at USD was almost identical to the one used at LBNL~\cite{meng2019fabrication}, with only minor adjustments to accommodate for different devices.

\subsection{Detector Characterization in Vacuum}
\begin{figure}[htbp] \centering
  \includegraphics[width=0.8\linewidth]{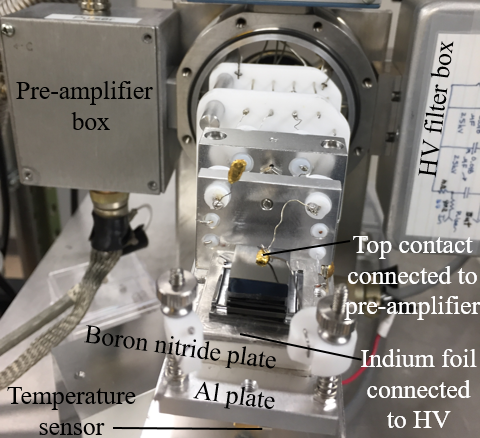}
  \caption{Internal structure of the vacuum cryostat at USD.}
  \label{f:setupUSD}
\end{figure}

Prior to the deployment of the detectors at MPI, their leakage currents, depletion voltages, and energy resolutions were measured in a vacuum cryostat at USD. Its internal structure is shown in Fig.~\ref{f:setupUSD}. The aluminum stage where the detectors were placed was cooled by a stainless steel tube filled with LN$_2$. A temperature sensor was placed at the bottom of the stage. The lowest operation temperature of the stage was measured to be 78~K. All the measurements were done one hour after the stage reached 78~K to allow the detector to be in equilibrium with the stage.

\begin{figure}[htbp]
  \includegraphics[width=\linewidth]{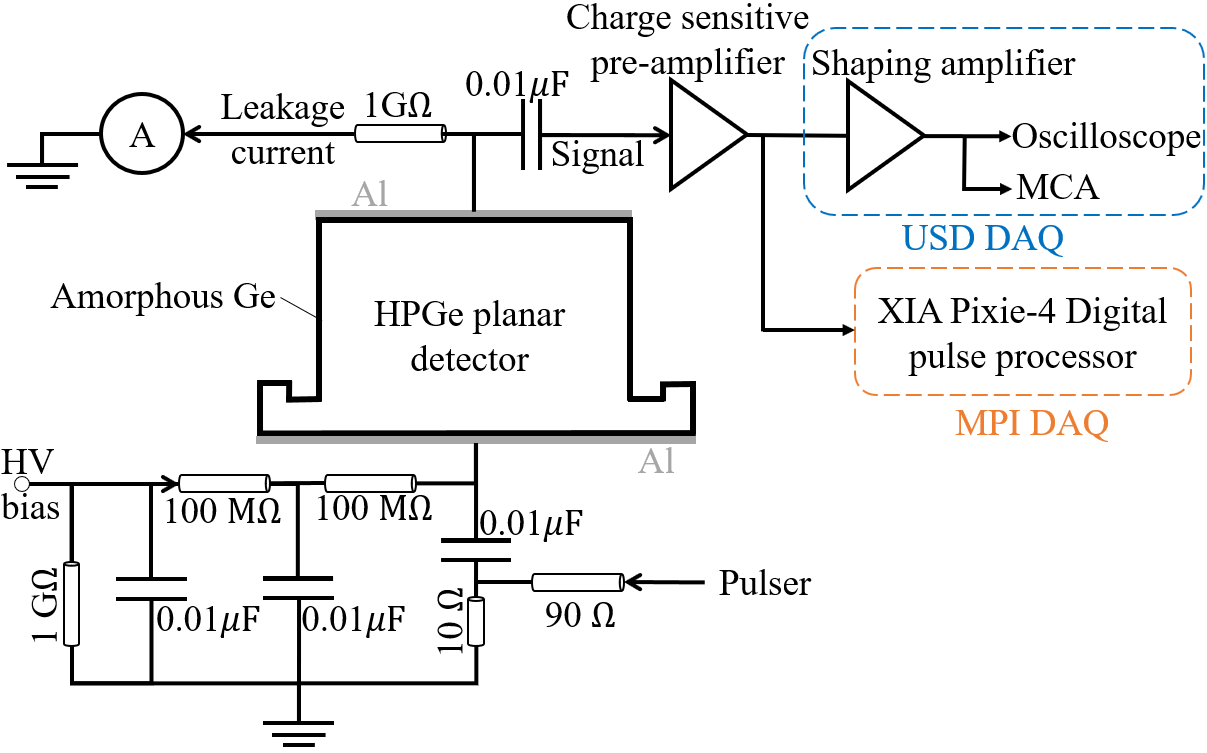}
  \caption{Electronic circuit for detector characterization.}
  \label{f:circuit}
\end{figure}

A schematic of the electronic circuit is shown in Fig.~\ref{f:circuit}. The detector was biased through its bottom contact, and read out at the top contact. Being a direct current, the leakage current, $I$, could not pass the 0.01~$\mu$F capacitor before the charge sensitive pre-amplifier, but the \SI{1}{\giga\ohm} resistor before the ammeter, and was measured there. Transient signals, however, could not pass the resistor, but the capacitor, and were amplified thereafter. The ammeter in the MPI setup was a Keithley picoammeter, which can measure leakage currents down to 20~fA. The instrument used at USD was a combination of a transimpedance amplifier and a regular multimeter, the precision of which was only 1~pA. The instruments had built-in noise-cancelling mechanisms. The displayed values were averages of a certain number of internal measurements. The leakage currents of the detectors in different environments at 1200~V are summarized in Table~\ref{t:det}.

Leakage currents in these mini planar HPGe detectors arise mainly due to charge injections at their top and bottom contacts (bulk leakage) and the leakage through defects on the side surfaces (surface leakage). The two types of leakages could be measured separately in one case as a ring of aluminum was etched away from the top contact, which was separated into a small central contact and a surrounding guard contact consequently. The leakage current measured from the central contact was mainly due to charge injections. The one measured from the guard contact was mainly due to the side surface leakage. Detector USD-R02 was used for such measurements before being repurposed for this study, hence it had two contacts on its top surface as shown in Fig.~\ref{f:R02}. Without such a structure, the measured leakage currents of the other two detectors were a sum of the bulk and the surface leakages.

The amount of leakage current is an important indicator for the quality of the contacts and side passivation of a detector. It increases with the bias voltage and the temperature, as predicted by the model developed by D\"ohler, Brodsky~\cite{dohl74, brod75, brodsky75} and Schottky~\cite{sze81} and successfully applied to amorphous germanium contacts on HPGe detectors~\cite{hall}. It may also change over the first few thermal cycles after fabrication and gradually stabilizes afterward\cite{looker2015leakage}. A detailed study of the leakage currents of the USD detectors can be found in Ref.~\cite{barrier}. On average, the bulk leakage is around a few pA, the surface leakage is around a few tens of pA, at 78~K. In contrast, detectors made at LBNL using USD crystals typically have a combined leakage below 1~pA.

To avoid charge trapping due to low electric field in some part of a detector, the operation voltage should normally be much higher than the depletion voltage. It is hence of interest to measure the latter to help determine the former. In addition, the depletion voltage of a planar detector, $V_d$, is associated with the net impurity concentration of the HPGe crystal through the following equation:
\begin{equation}\label{e:imp}
  |N_A-N_D|=2\varepsilon V_d/e/D^2,
\end{equation}
where $N_A$ and $N_D$ are the $p$ and $n$-type impurity concentrations, respectively, $\varepsilon$ is the permittivity of Ge, $e$ is the elementary charge, and $D$ is the detector thickness. The measurement of $V_d$ can then be used to verify the net impurity level given by the Hall-effect measurement of the crystal.

A scan of the detector capacitance, $C_d$, at various bias voltages, $V_b$, can be used to determine $V_d$. This can be understood as follows. As the bias voltage of the detector, $V_b$, goes up, the thickness of the depleted region, $d$, increases, the detector capacitance, $C_d$, goes down, because $C_d$ is anti-proportional to $d$. When the detector is fully depleted, $d=D$, and cannot increases any more, $C_d$ becomes a constant thereafter. The bias voltage at the point where the $C_d$-$V_b$ curve starts to flatten out is therefore the depletion voltage, $V_d$.

\begin{figure}[htbp]
    \includegraphics[width=\linewidth]{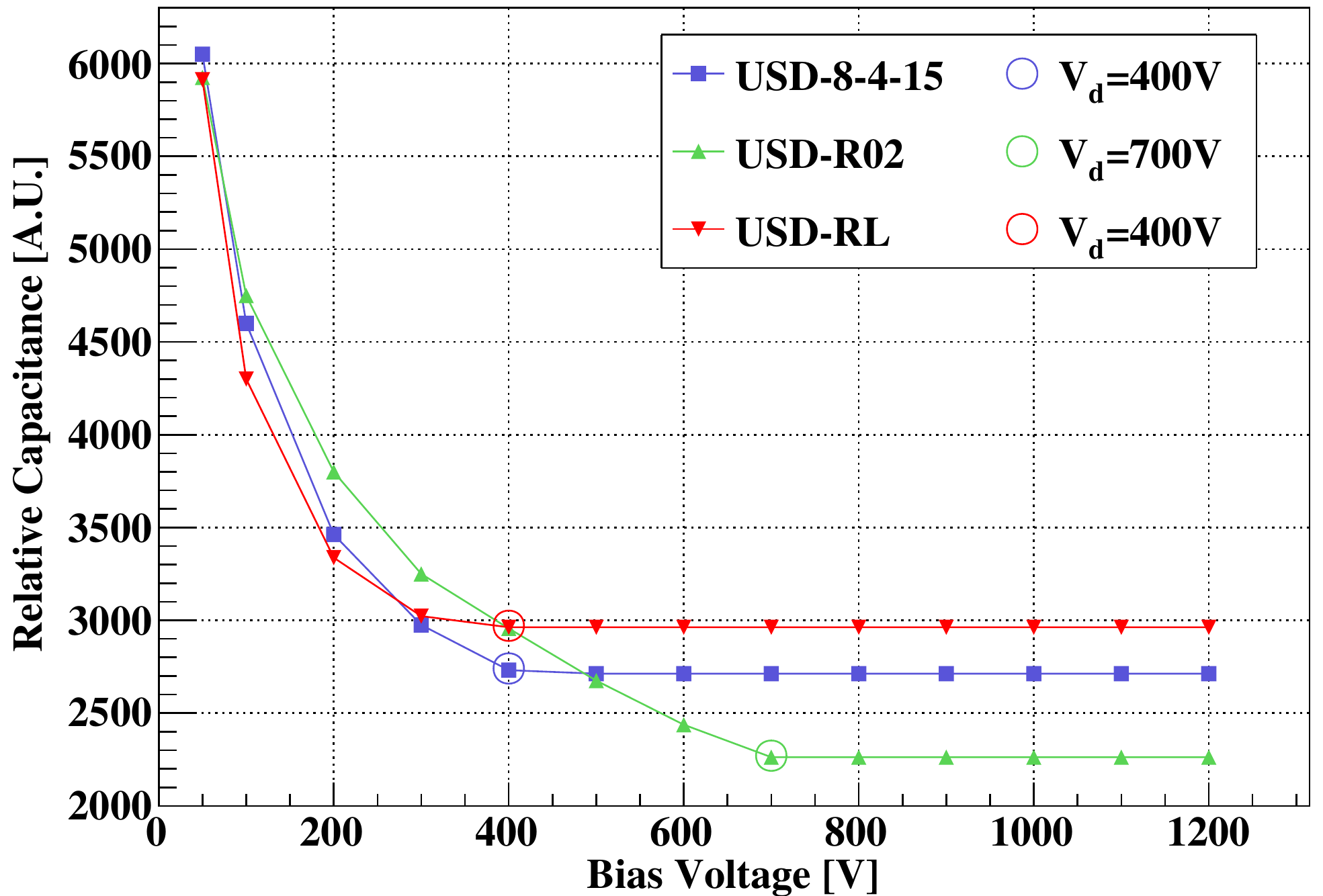}
    \caption{Relative capacitance as a function of bias voltage.}
    \label{f:CV}
\end{figure}

$C_d$ was not measured directly, its bias voltage dependence was estimated as follows:
\begin{itemize}
  \item Inject step voltage pulses with a fixed amplitude, $V_p$, from a pulser to the circuit.
  \item The voltage change is converted to charge injection to the detector through the $0.01~\mu$F capacitor in between the pulser and the detector.
  \item This change of charges can be converted to a voltage pulse, $V_o$, by the charge-sensitive pre-amplifier.
  \item Given a fixed charge injection, $q$, the output voltage, $V_o$, is proportional to $C_d$, according to the relation $q = CV$ that applies to an ideal planar capacitor.
\end{itemize}
The $V_o$--$V_b$ curve hence has the same behavior as the $C_d$--$V_b$ curve. Thus, the full depletion voltage can be determined using the former.  Figure~\ref{f:CV} shows the $V_o$--$V_b$ (relative capacitance versus bias) curves for the three detectors. Their bias voltages are marked in the figure and summarized in Table~\ref{t:det}.  The impurity concentrations of individual crystal were calculated using Eq.~\ref{e:imp} with the measured values of $V_d$. They are listed in Table~\ref{t:det}.

The energy resolution of an HPGe detector is a convolution of three major components, the electronic noise $\Delta E_e$, the fluctuation of the number of charge carriers in their creation process $\Delta E_n$, and a component due to trapping of charge carriers $\Delta E_t$: $$\Delta E^2 = \Delta E^2_e +\Delta E^2_n+\Delta E^2_t.$$
Since pulser signals do not originate from physical events, the fluctuation of their pulse heights depends only on the electronic noise. The resolution of the peak in the energy spectrum due to pulse-injection is hence a good indicator of the electronic noise $\Delta E_e$. A $\gamma$-ray peak should be wider than the pulser peak due to the additional contributions of $\Delta E_n$ and $\Delta E_t$.

\begin{figure}[htbp]
  \includegraphics[width=\linewidth]{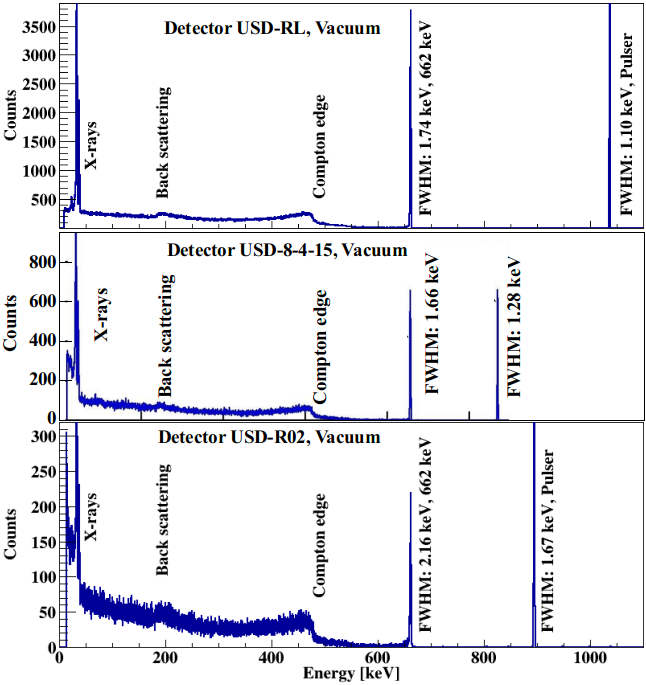}
  \caption{Energy spectra obtained with the LBNL vacuum cryostat and a $^{137}$Cs source outside of the cryostat.}
\label{f:specVac}
\end{figure}

This is shown clearly in Figure~\ref{f:specVac}, the energy spectra measured when the detectors were biased at 1200~V in the LBNL vacuum cryostat. They were taken with a $^{137}$Cs radioactive source placed outside the cryostat above the detector. Rectangular pulses with a fixed amplitude were used to generate the pulser peak above 662~keV. The resolution of the $\gamma$-ray peak, $\Delta E$, was always slightly larger than that of the pulser peak, $\Delta E_e$.

\subsection{Cryostat at MPI}
\begin{figure}[htbp]
  \includegraphics[width=\linewidth]{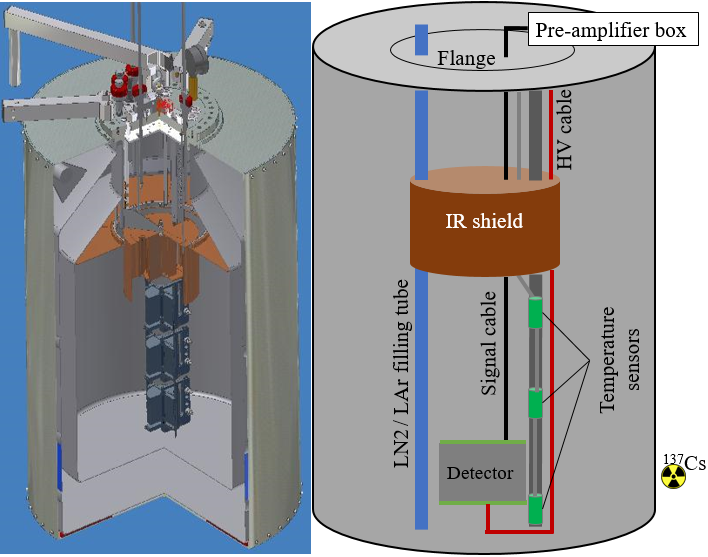}
  \caption{Left: technical drawing of the MPI cryostat. Right: schematics of its internal wiring.}
  \label{f:setupMPI}
\end{figure}

A liquid argon cryostat named Gerdalinchen II was developed by the germanium detector group at MPI for the operation of up to three segmented HPGe detectors in cryogenic liquids~\cite{gerdalinchen}. An artist view is shown in the left part of Fig.~\ref{f:setupMPI}. It was used for the operation of USD detectors in LN$_2$ and LAr. The top flange of Gerdalinchen II is opened vertically for installation. Detector holders and the central part of the infrared (IR) shield are attached to a vertical stainless steel bar, which is fixed to the top flange. The assembly is lifted together with the top flange.

\begin{figure}[htbp] \centering
  \includegraphics[width=0.6\linewidth]{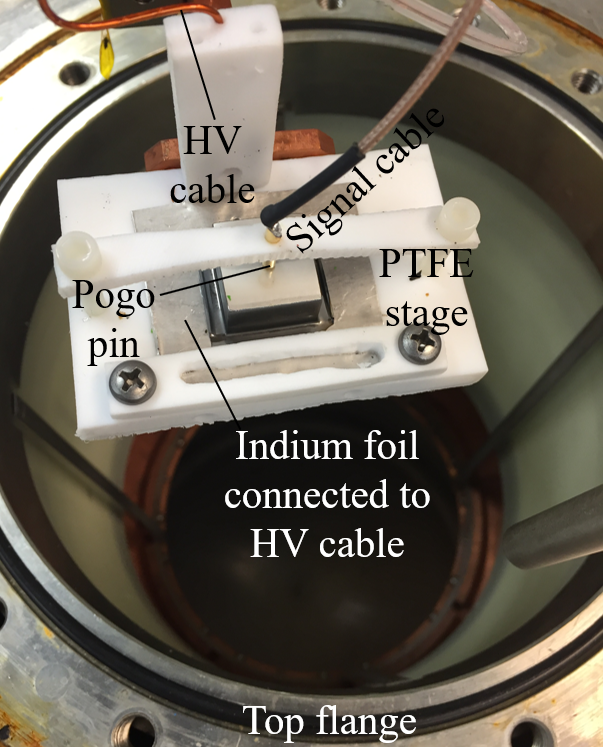}
  \caption{Detector to be lowered into the MPI cryostat.}
  \label{f:GerdalinchenII}
\end{figure}

For the operation of the USD detectors, a simple PTFE stage was mounted to the lowest position on the vertical bar as shown in Fig.~\ref{f:GerdalinchenII}. An indium foil was pressed on top of the stage using two PTFE bars. A rigid high voltage (HV) cable went through the vertical PTFE bar and was pushed tightly against the indium foil to provide the bias voltage. The detector was placed on top of the indium foil. A pogo pin connected to the signal cable was pressed lightly on the top surface of the detector. Three PT100 temperature sensors were mounted along the stainless steel bar. The lowest one was slightly below the bottom of the detector. The middle one was a few centimeters above the detector. The top one was close to the IR shield. They were used to monitor the liquid level in the cryostat. The internal wiring scheme is shown on the right side of Fig.~\ref{f:setupMPI}.

There were safe procedures to fill and empty Gerdalinchen II to avoid any frosting of the detectors.

\subsection{Detector Operation in Liquid Nitrogen}
\label{s:ln2}

The detectors were first operated in LN$_2$. The same measurements as those done at USD were repeated in the new environment: the leakage current and the relative capacitance as functions of bias voltage, and the energy resolution of the 662~keV $\gamma$-ray peak from a collimated 5~MBq $^{137}$Cs source at 1200~V.

\begin{figure}[htbp]
  \includegraphics[width=\linewidth]{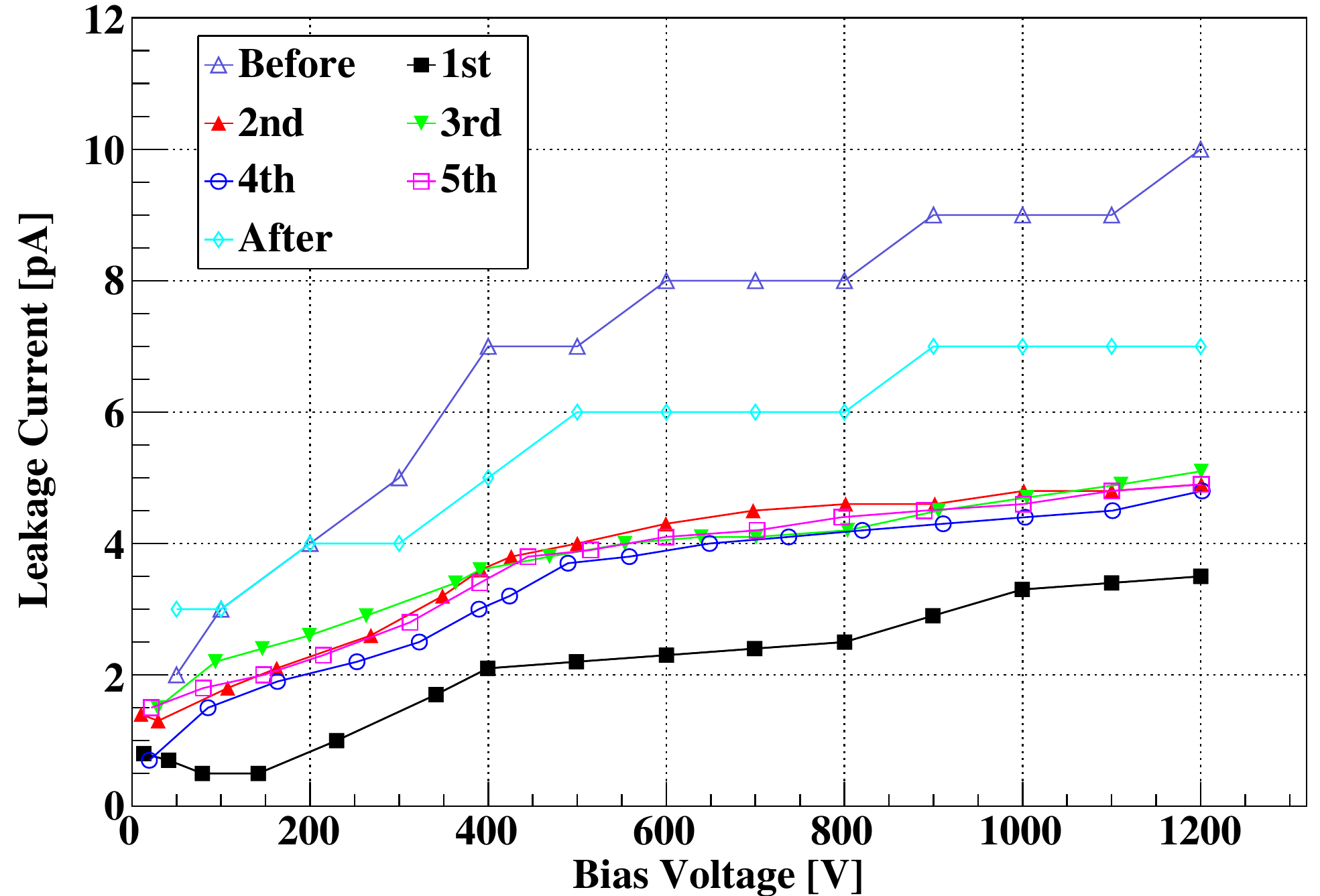}
  \caption{Leakage current of detector USD-RL as a function of its bias voltage in LN$_2$, except for the ``Before'' and ``After'' data sets, which were measured in vacuum at USD before and after the MPI deployment, respectively. The numbers denote thermal cycles in LN$_2$.}
  \label{f:LRLN2}
\end{figure}

\begin{figure}[htbp]
  \includegraphics[width=\linewidth]{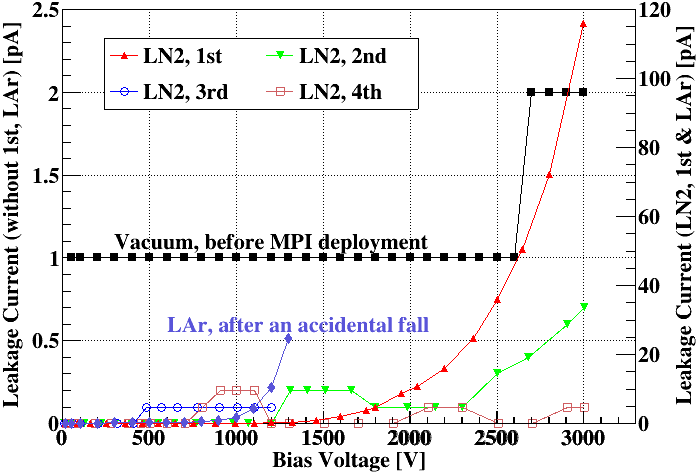}
  \caption{Leakage current of detector USD-8-4-15 as a function of its bias voltage in various environments. The numbers in the legend denote thermal cycles in LN$_2$. The scale for the first cycle in LN$_2$ and the LAr measurement is on the right.}
  \label{f:MarkLN}
\end{figure}

\begin{figure}[htbp]
\includegraphics[width=\linewidth]{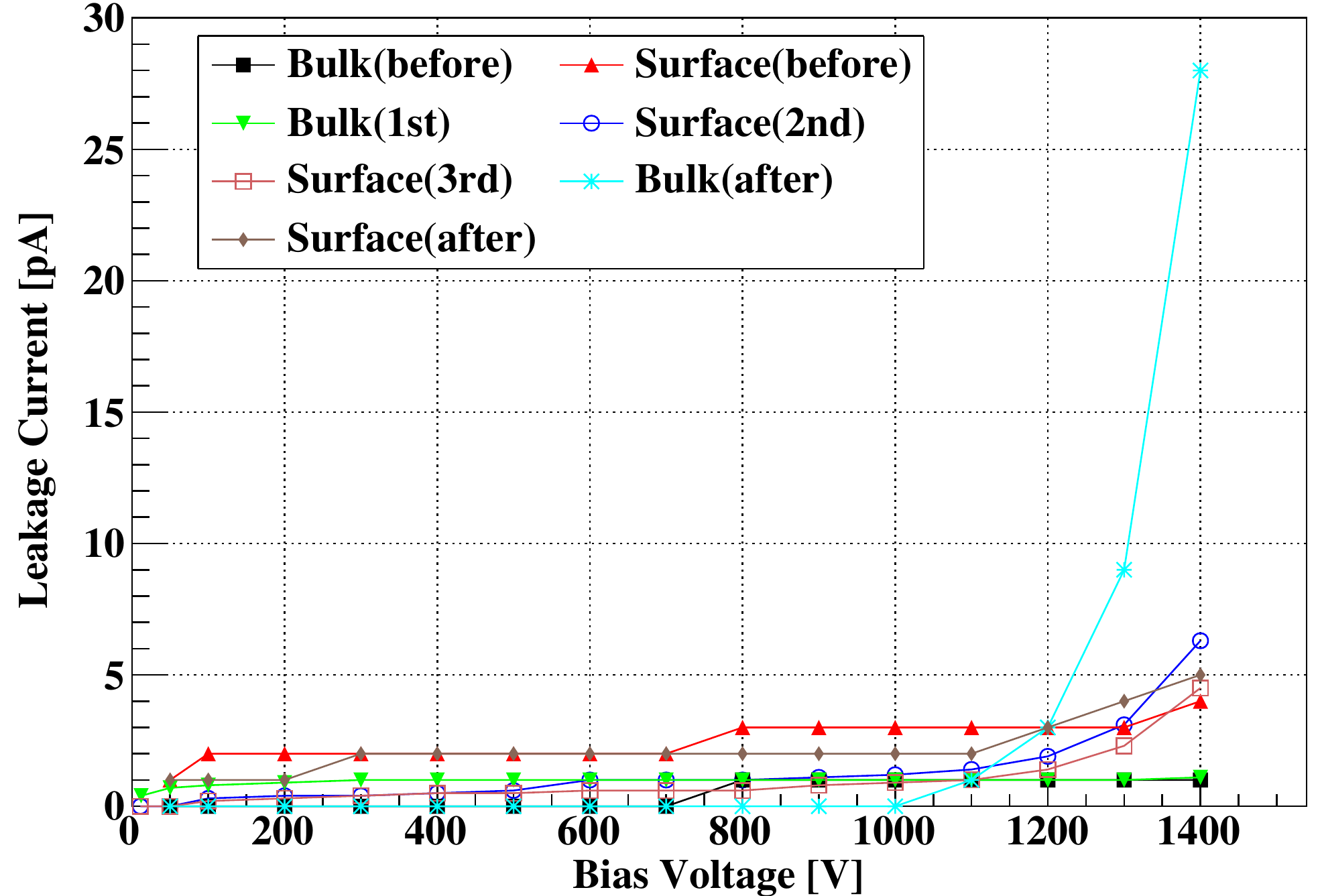}
  \caption{Leakage currents of detector USD-R02 versus its bias voltage in LN$_2$, except for the data sets marked with ``before'' and ''after'', which were measured in the vacuum cryostat at USD before and after the MPI deployment. The bulk leakage currents were measured through the central contact. The surface leakage currents were measured through the guard contact.}
  \label{f:R02LN2}
\end{figure}

Figs.~\ref{f:LRLN2}, \ref{f:MarkLN} and \ref{f:R02LN2} show the leakage currents of the three detectors as functions of their bias voltages after each thermal cycle in LN$_2$. For reference, data sets taken in the vacuum cryostat at USD before and after the MPI deployment were plotted in the same figures. Each data point was recorded a few tens of seconds after a new bias voltage was applied, when the reading stabilized.

The leakage current of detector USD-RL measured during the first cooling cycle was 3.5~pA at 1200~V, shown as the last point in the lowest curve in Fig.~\ref{f:LRLN2}. It was monitored thereafter and a slow steady increase was observed over time. After about an hour, the leakage current stabilized at 5.1~pA. The leakage current after that was very stable over five thermal cycles. A current of 5~pA at 1200~V was always observed.  Such a slow increase of the leakage current was not observed in other detectors in these studies. It might be due to a gradual development of a small leakage channel on the side surface of the detector.  The data sets denoted as ``before'' and ``after'' were measured in the vacuum cryostat at USD before and after the MPI deployment. They are slightly higher than those measured in LN$_2$. This is because the real temperature of the detectors in the vacuum cryostat was a few degrees higher than the LN$_2$ temperature, and the leakage current increases with temperature~\cite{brodsky75, sze81}. Overall, there was no significant change of the leakage current for detector USD-RL measured in different thermal cycles and environments, and all currents were below 10~pA up to 1200~V.

As shown in Fig.~\ref{f:MarkLN}, the leakage current of detector USD-8-4-15 was basically around 1~pA in both environments, except for the data set measured during the first cool down, which increases rapidly after 1500~V. One possible explanation is that some dust attached itself to the surface of this detector during the process of moving it from USD to MPI, and created a surface leakage channel, which was washed or blown off from the surface in the first cooling cycle; as the leakage channel was removed, the detector behaved normally afterward.

\begin{figure}[htbp] \centering
  \includegraphics[width=\linewidth]{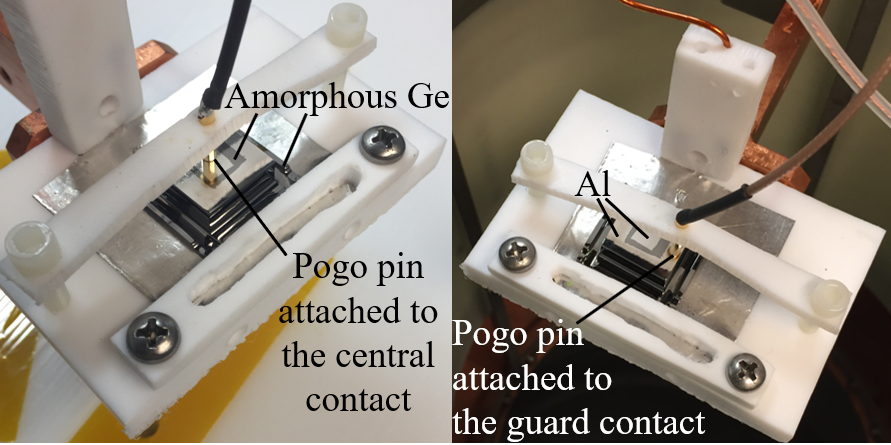}
  \caption{Two different contact schemes of the guard-ring detector USD-R02 in the MPI cryostat.}
  \label{f:R02}
\end{figure}

Only one read-out channel could be used in the MPI cryostat. The central and the guard contacts on the top surface of detector USD-R02 were connected to it through a pogo pin one at a time, the other contact was left floating as shown in Fig.~\ref{f:R02}. In contrast, both contacts were read out in the vacuum cryostat at USD.

As shown in Fig.~\ref{f:R02LN2}, the leakage currents of USD-R02 in different contacts, environments and thermal cycles were mostly below 5~pA, except for the bulk leakage measured at USD after the MPI deployment, which may be due to a damage to the detector surface during the shipment as a small scratch was observed on its top surface.

The ``capacitance'' versus bias voltage curves measured in LN$_2$ were basically identical to those measured in vacuum. The depletion voltages determined this way were the same as those determined at USD. This was as expected since the depletion voltage is basically determined by the impurity level of the crystal and should not change with the environment at a given temperature.

\begin{figure}[htbp]\centering
  \includegraphics[trim=0 0 0 34,clip,width=\linewidth]{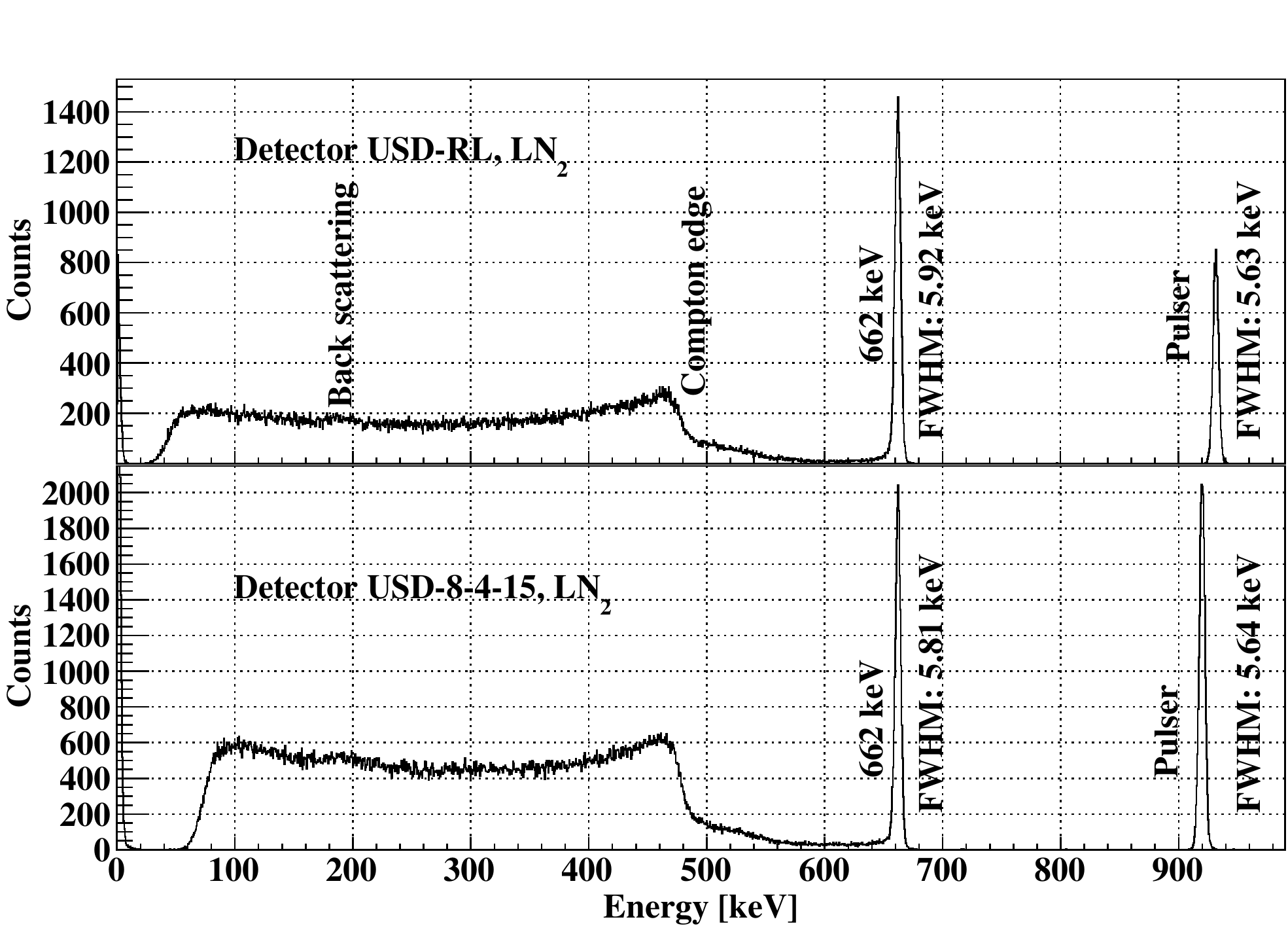}
  \caption{Energy spectra of $^{137}$Cs taken in LN$_2$. No spectrum was taken with USD-R02 since the source was temporarily unavailable for the measurement.}\label{f:specLN2}
\end{figure}

The energy spectra of $^{137}$Cs taken with the detectors in LN$_2$ is shown in Fig.~\ref{f:specLN2}. The FWHMs of the pulse peaks were about 5.6~keV. Due to the large noise, no quantitative statement can be made regarding the influence of cryogenic liquids on the energy resolution of these detectors. Nevertheless, the spectra measured in LN$_2$ were very similar to those measured in vacuum shown in Fig.~\ref{f:specVac}, which proved that they worked as spectroscopic devices in LN$_2$.

There was no effort made to optimize the read-out as it was beyond the goal of this initial study. A standard way to improve this is to move the front-end jFET from the pre-amplifier board to somewhere inside the cryostat, a few centimeters above the liquid level, to reduce the signal cable length and to achieve an optimized operating temperature of the jFET. This and other measures will be taken in the future to reduced the impact of the electronic noise.

\subsection{Detector Operation in Liquid Argon}

\begin{figure}[htbp]
  \includegraphics[width=\linewidth]{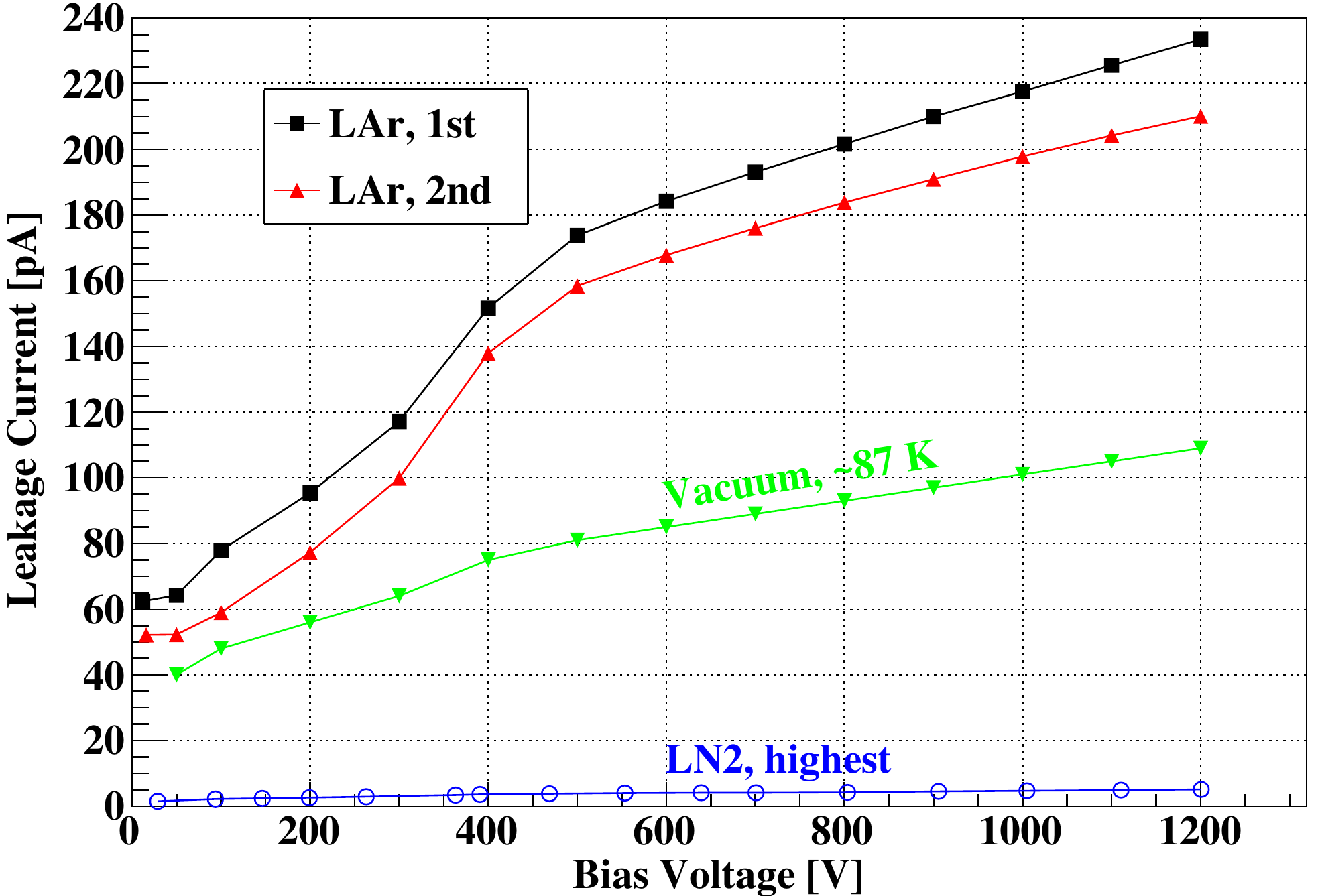}
  \caption{Leakage currents of detector USD-RL versus its bias voltage in LAr.  The numbers denote individual thermal cycles. Also plotted are the highest leakage current measured with the same detector in LN$_2$ and the one measured in the vacuum cryostat at USD after its deployment at MPI.}
  \label{f:LRLAr}
\end{figure}

\begin{figure}[htbp]
  \includegraphics[width=\linewidth]{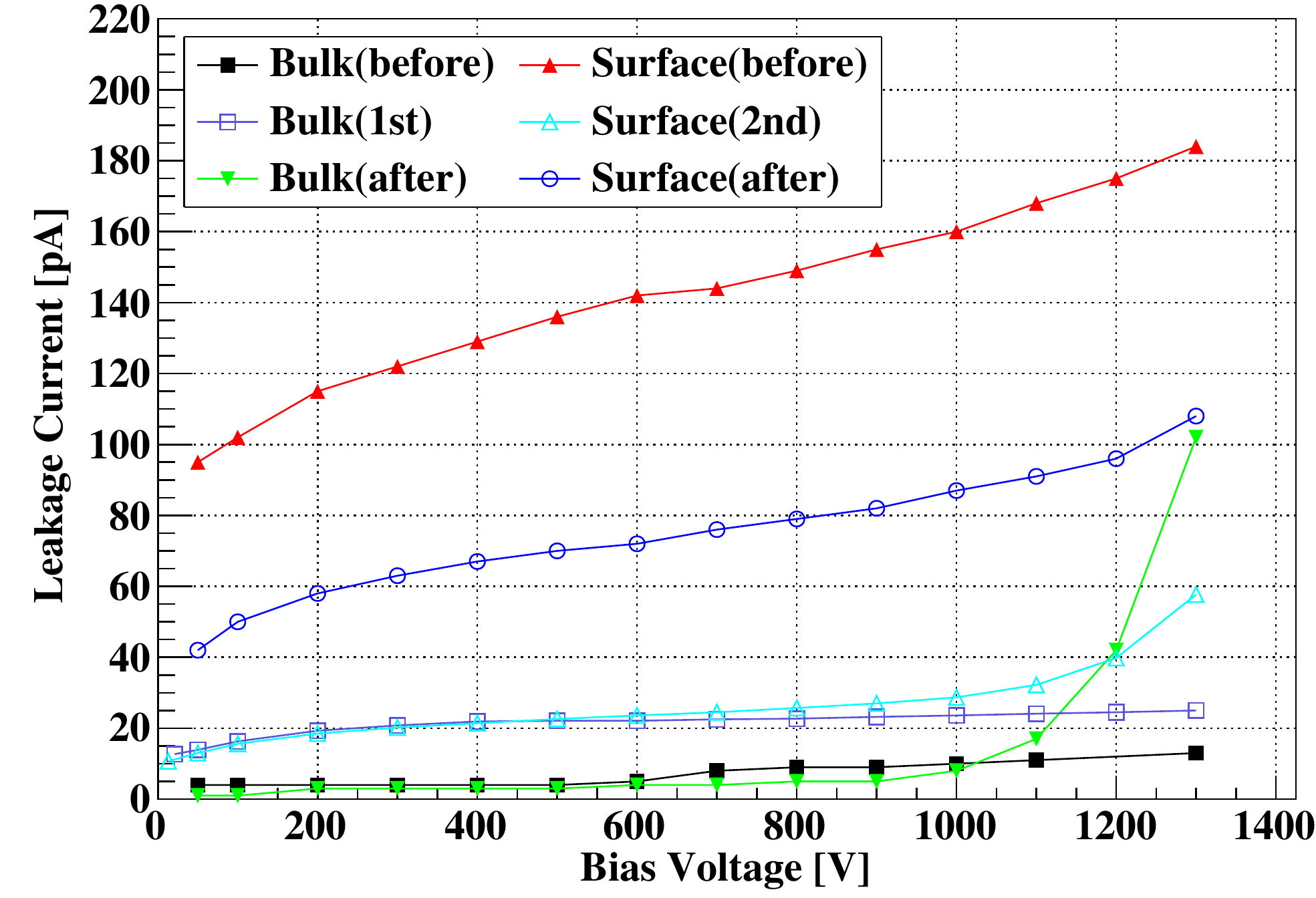}
  \caption{Leakage currents of detector USD-R02 versus its bias voltages in LAr, except for the ones labeled ``before'' and ``after'', which were measured in the USD vacuum cryostat at 90~K. The numbers denote the thermal cycles in LAr. Bulk leakage currents were measured through the central contact, surface ones were through the guard contact.}
  \label{f:R02LAr}
\end{figure}

The same measurements were repeated with the same detectors in LAr using the same cryostat at the MPI.  Figs.~\ref{f:MarkLN}, \ref{f:LRLAr} and \ref{f:R02LAr} show the leakage currents of the three detectors as functions of their bias voltages after each thermal cycle in LAr. For reference, data sets taken at 90~K in the vacuum cryostat at USD before and after the MPI deployment were plotted in the same figures, and labeled as ``before'' and ``after'', respectively.

Detector USD-RL went through two more thermal cycles in LAr. The leakage currents were about 20 times higher than those measured in LN$_2$.

Detector USD-8-4-15 was operated in LAr once.  Below 800~V, the leakage current was below 1~pA. Its significantly lower leakage current is a clear evidence that the quality of the amorphous germanium surface made at LBNL~\cite{looker2015leakage, amman2018optimization} was better than that made at USD~\cite{meng2019fabrication, wei2018investigation}. The quick rise of the leakage current above 800~V was due to damage to the detector when it fell from the PTFE stage during the preparation of the fifth thermal cycle in LN$_2$. Nonetheless, it still had the best performance compared to the other two detectors.

USD-R02 was operated twice in LAr, the first time with its central contact connected to the signal cable, the second time with its guard contact connected to the signal cable.  The bulk leakage increased a few times compared to those in LN$_2$, the surface leakage increased about 20 times. Note, that the leakage current of detector USD-RL in LAr was also increased by about 20 times, which was probably also dominated by surface leakage.

The measurements at USD after the MPI deployment were done at about 90~K instead of 78~K to be closer to the LAr temperature. The leakage current of the central contact of USD-R02 (triangle data points connected with green lines) rose quickly after 1,100 V, which might be due to a damage to the detector top surface during the shipment back to USD as a small scratch was observed there.

\begin{figure}[htpb]
  \includegraphics[width=\linewidth]{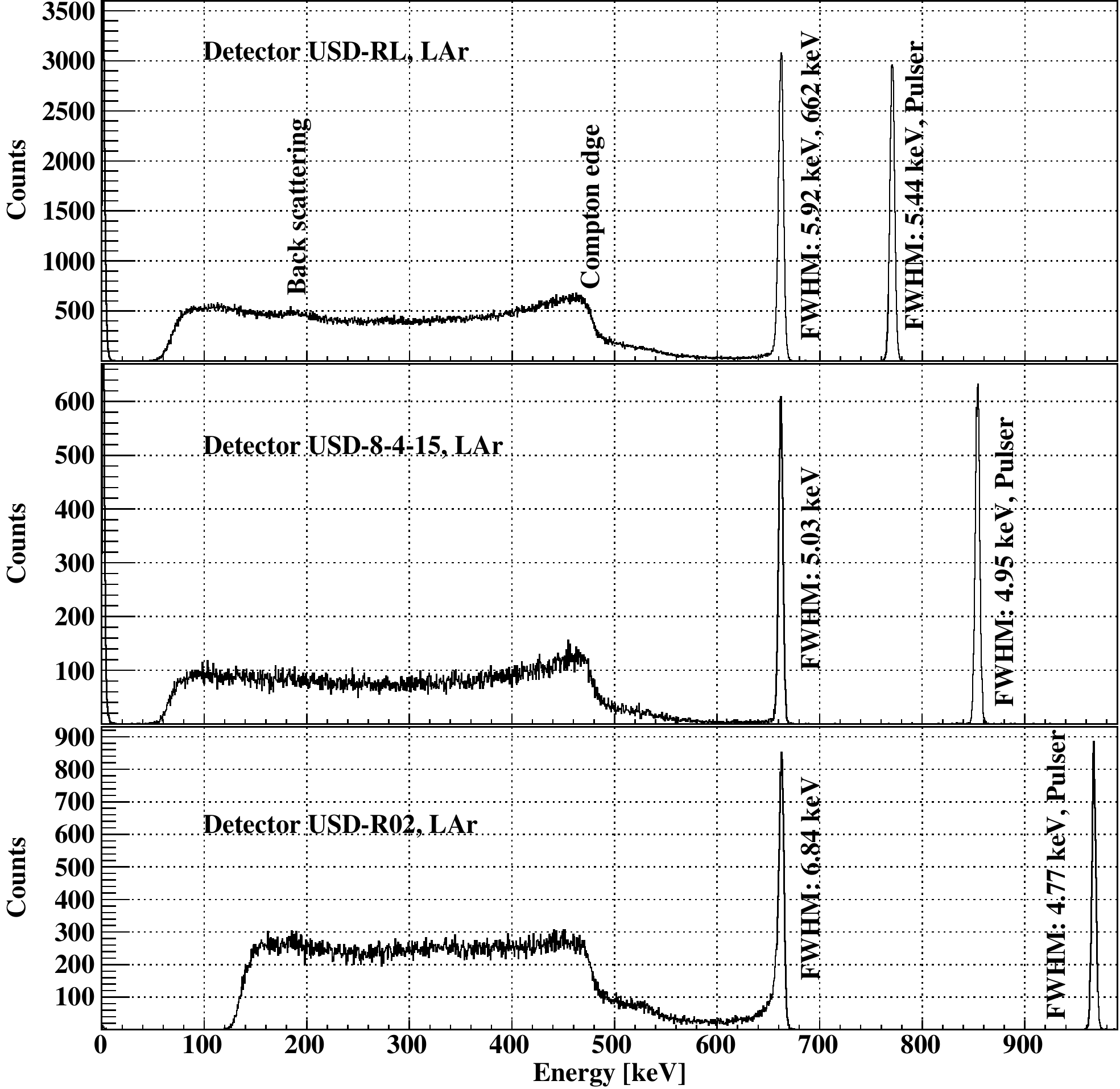}
  \caption{Energy spectra of $^{137}$Cs taken in LAr.}
  \label{f:specLAr}
\end{figure}

The energy spectrum of $^{137}$Cs measured with detector USD-RL biased at 1200 V in LAr is shown in Fig.~\ref{f:specLAr}. The energy resolution and the noise level were similar to those measured in LN$_2$, despite of much larger leakage currents in LAr than those in LN$_2$, which suggested the dominating contribution to the noise from the read-out system.

\subsection{Characterization in Vacuum Again}

Detector USD-RL and USD-R02 were characterized in the vacuum cryostat again after their operations in cryogenic liquids, which confirmed that the detectors could still function normally after the deployment at MPI. Their leakage current measurement results were shown together with those measured in LN$_2$ and LAr in Fig.~\ref{f:LRLN2}, \ref{f:R02LN2}, \ref{f:LRLAr} and \ref{f:R02LAr} as references. The energy spectrum of $^{137}$Cs taken with detector USD-RL at 1200~V, 78~K in vacuum is shown in Fig.~\ref{f:specVac}. Detector USD-8-4-15 was left at MPI for future investigations. No measurement with this detector was repeated in the vacuum cryostat at USD.

\section{Cross Comparison}

\subsection{Different Detectors in Same Environment}

\begin{figure}[htpb]
  \includegraphics[width=\linewidth]{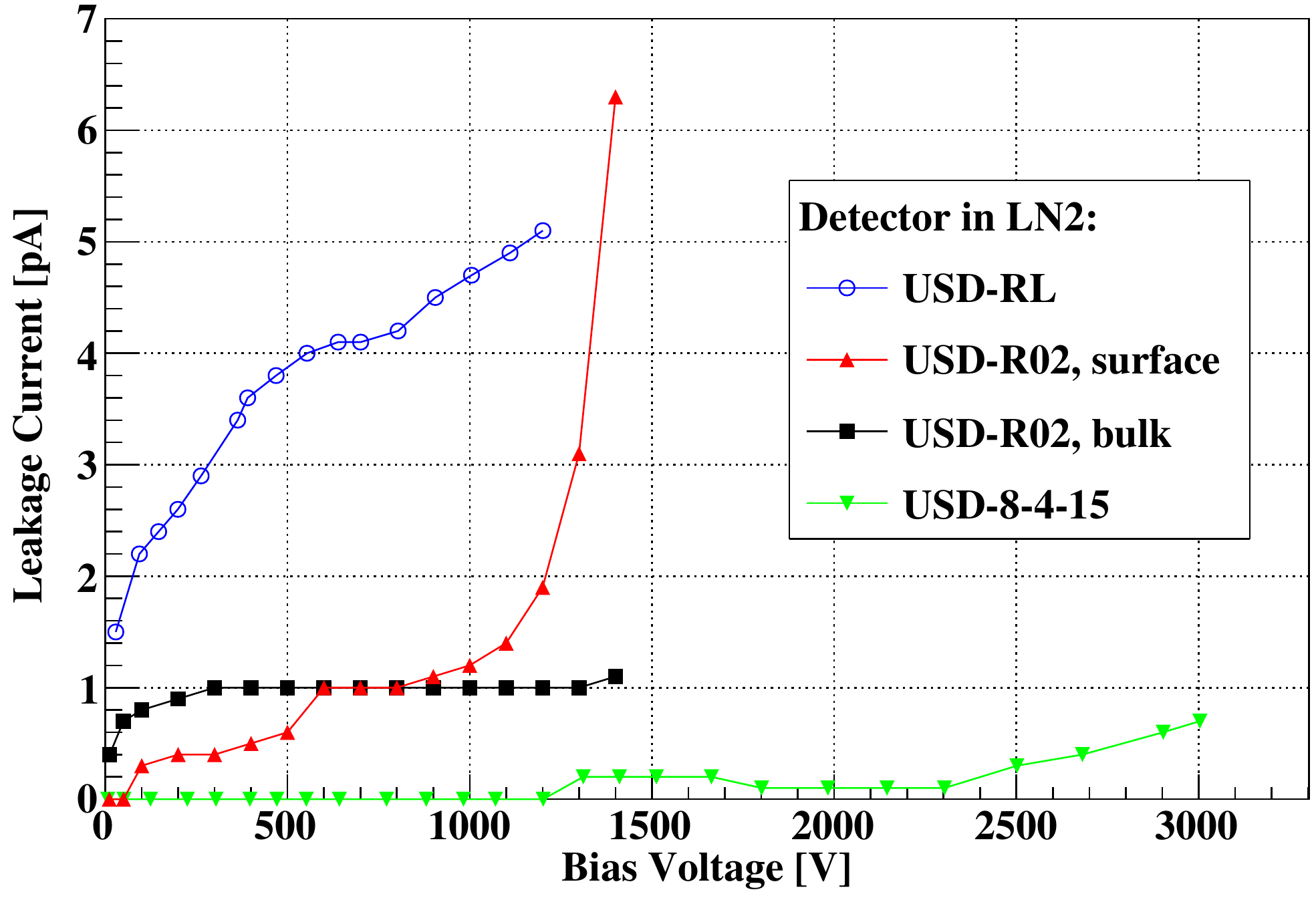}
  \caption{Highest leakage currents of the three detectors measured in LN$_2$.}
  \label{f:lcn2}
\end{figure}

\begin{figure}[htpb]
  \includegraphics[width=\linewidth]{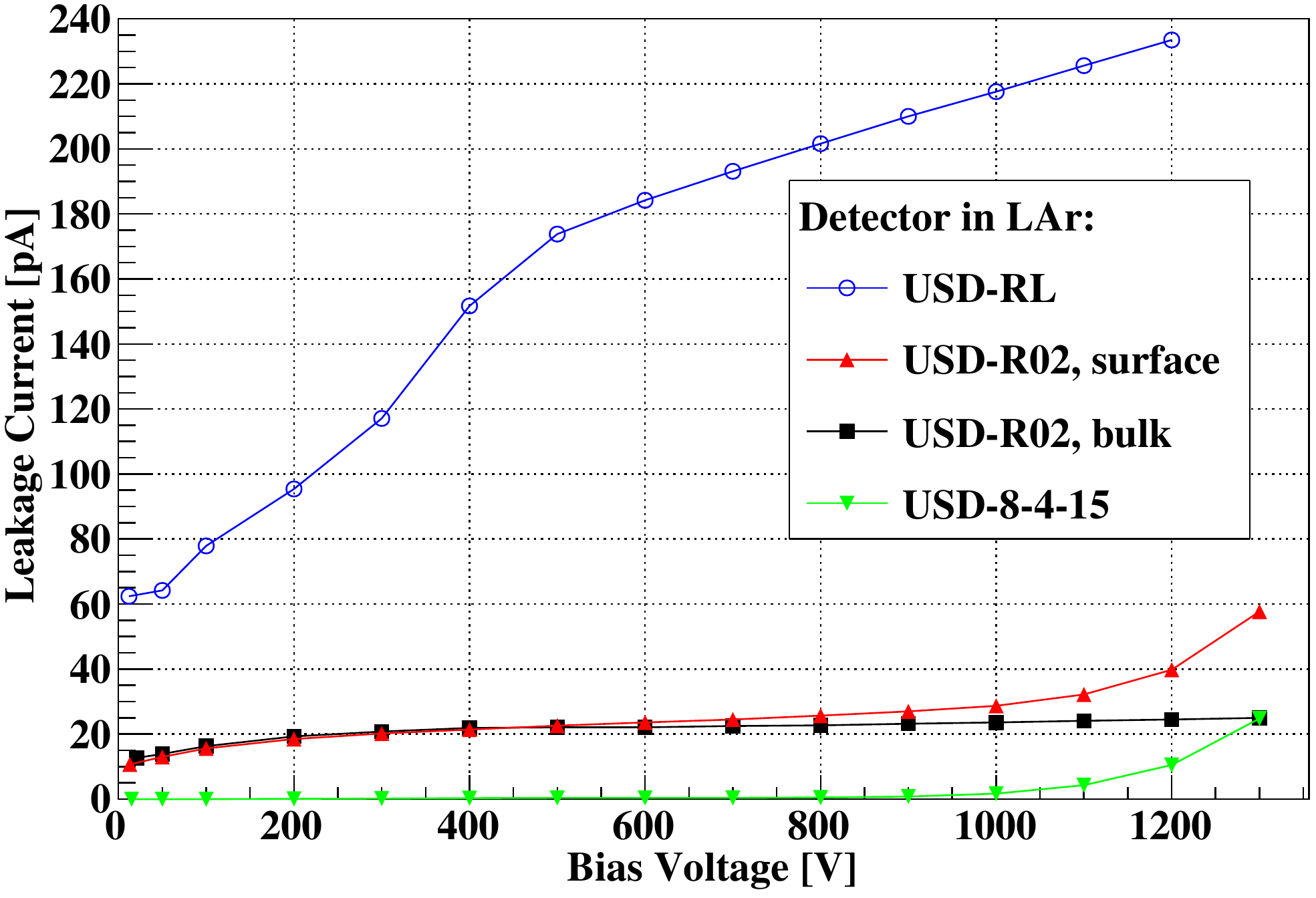}
  \caption{Highest leakage currents of the three detectors measured in LAr.}
  \label{f:lcar}
\end{figure}

Fig.~\ref{f:lcn2} and \ref{f:lcar} compare the leakage currents of the three detectors in LN$_2$ and LAr, respectively. USD-RL exhibited the highest leakage current among them in both environments, while USD-8-4-15 exhibited the lowest among all. The side surface leakage currents of USD-R02 were typically higher than its bulk leakage currents through the central contact around operational voltages in both environments. These results are consistent with more thorough investigations done in vacuum at USD with more sample detectors~\cite{meng2019fabrication, wei2018investigation}, that is, the performance of the detectors made at USD has yet to be improved to match that of the detectors made at LBNL by Mark Amman, in particular, the quality of the side surface. Nevertheless, the performance of USD-8-4-15 in both cryogenic liquid is very encouraging.

\subsection{Same Detector in Different Environments}
\begin{figure}[htpb]
  \includegraphics[width=\linewidth]{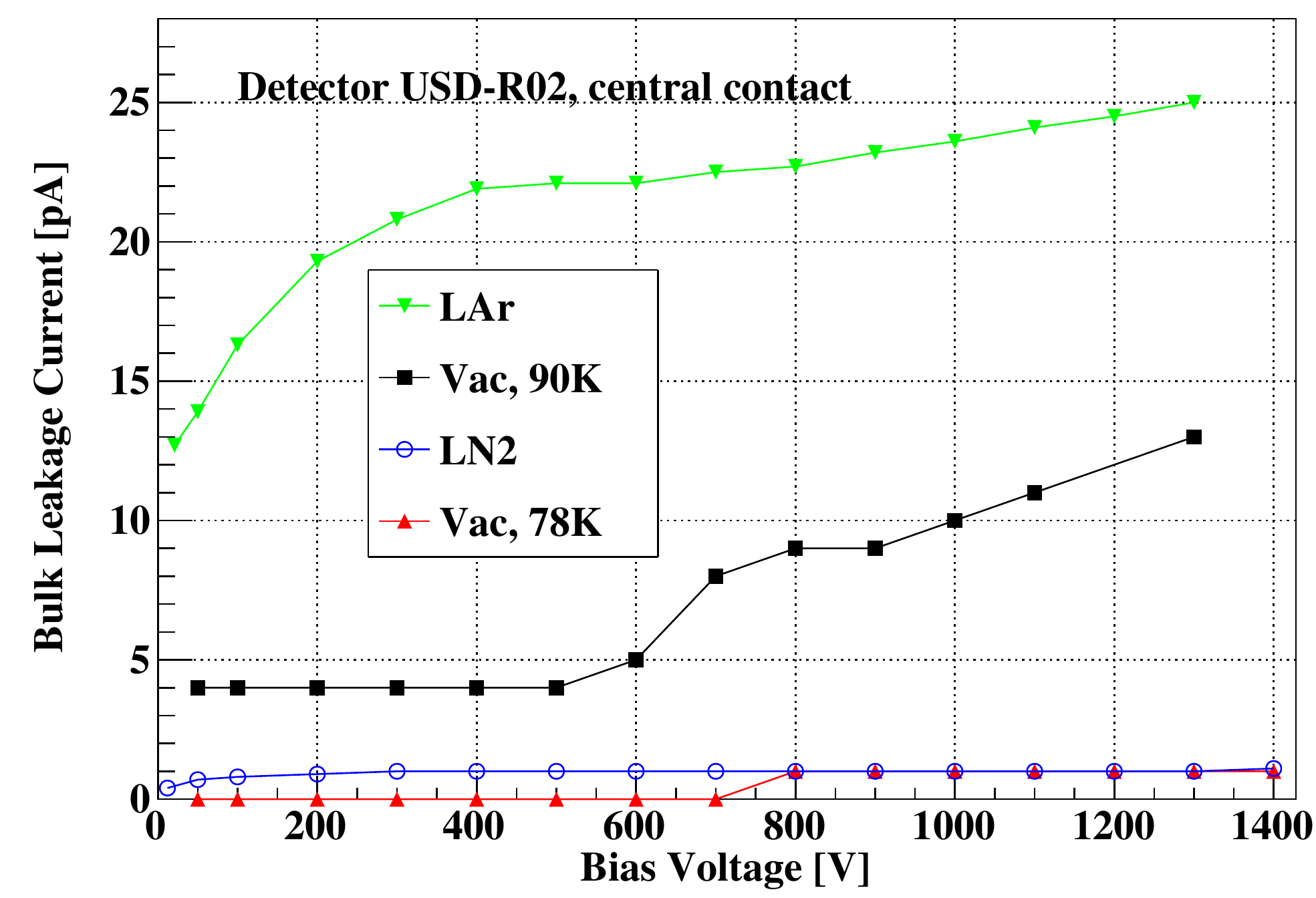}
  \caption{Largest leakage currents of detector USD-R02 in various environments through its central contact.}
  \label{f:r2b}
\end{figure}

Fig.~\ref{f:r2b} compares the bulk leakage currents through the central contact of USD-R02 measured in various environments. The contribution of side surface leakage was minimized in those measurements. Leakage currents measured at higher temperatures were higher than those at lower temperatures. Such a temperature dependence is well documented in the literature~\cite{dohl74, brod75, brodsky75, sze81, hall, looker2015leakage,  barrier}.

The difference between the LAr and vacuum measurements at similar temperatures may have two possible explanations. First, LAr may have some negative impact on the charge-carrier blocking capability of the amorphous germanium contact. Second, more time is needed for the surface property of USD-R02 to stabilize, as the slow decreasing of leakage currents over shelf time has been observed in LBNL detectors as well~\cite{looker2015leakage}. Similar measurements with the same detector need to be repeated a few times with some time intervals in between to exclude one of the possibilities.

The leakage current measured in vacuum at around 78~K seems lower than that measured in LN$_2$ below 800~V. However, they were measured with two different sets of equipment. Taking into account the precision of the equipment, they are consistent with each other.

Leakage currents of USD-8-4-15 in various environments are compared with each other in Fig.~\ref{f:MarkLN}. Excluding the impact of the accidental fall, they were all $\leq 1$~pA below 1200~V. The precision of the experimental setup was not enough to tell the subtle difference at that level.

\begin{figure}[htbp]\centering
  \includegraphics[width=\linewidth]{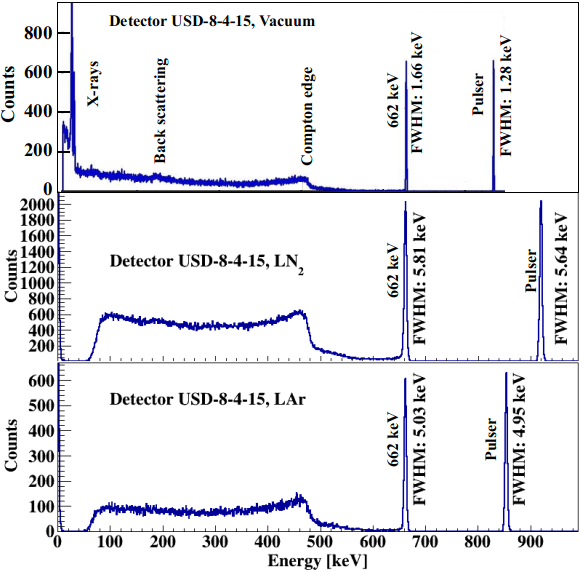}
  \caption{Energy spectra taken with detector USD-8-4-15 in various environments.}\label{f:smk}
\end{figure}

Energy spectra taken with USD-8-4-15 in various environments are plotted again in Fig.~\ref{f:smk} for easy comparison. The much wider pulser peaks in LN$_2$ and LAr compared to that in vacuum clearly indicate the dominating contribution of the electronic noise from the read-out system to the overall energy resolution of the $\gamma$-ray peaks. The large noise prevented a meaningful extraction of the intrinsic resolution of the detector from these measurements as such an attempt would suffer from large uncertainty from the subtraction of two numbers close in their values.

A high energy threshold was set for the measurements in LN$_2$ and LAr to maintain a reasonable trigger rate. The $X$-ray lines from the $^{137}$Cs source hence could not be recorded. Other than that, main structures exhibited in these spectra are very similar to that taken in vacuum.

\section{Conclusion and Outlook}
The possibility of operating HPGe detectors with thin amorphous germanium contacts directly in LN$_2$ and LAr has been demonstrated experimentally for the first time. Three mini planar detectors with such contacts made at LBNL and USD using USD HPGe crystals survived long-distance transportation, multiple thermal cycles in both cryogenic liquids, and showed reasonable leakage currents and spectroscopic performance.  The leakage currents measured for the best detector were under 1~pA at bias voltages well above the depletion voltage. The leakage currents in LAr of the other two detectors were much higher than those measured in LN$_2$, mainly due to the side surface leakage.

Due to completely different geometric configurations of the tested detectors and PPC detectors used in $0\nu\beta\beta$ decay experiments, no direct comparison can be made between the leakage currents measured here in LAr and those measured with the detectors used in GERDA~\cite{heider2008performance, palioselitis2015experience}. The USD group is working on the fabrication of mini PPC detectors with their entire surfaces covered by amorphous germanium. Some initial results will be soon published in another paper. Long term operations of such detectors in the MPI setup will be carried out in the future to further verify the feasibility of such a technique for $0\nu\beta\beta$ decay experiments.

Furthermore, it has been observed by the GERDA collaboration that the leakage current through the passivated end surfaces of some of their detectors in LAr increased after long-term operation or irradiation with $\gamma$-ray sources~\cite{heider2008performance, palioselitis2015experience}. It is hence of interest to monitor the leakage current through the side surface of a planar detector passivated with amorphous germanium during long-term operation in LAr. Such measurements will be done with planar detectors with guard contacts at least a year after their fabrication to let their amorphous germanium surfaces stabilize prior to their operation in LAr.

In summary, thin amorphous germanium contacts passed some preliminary survivability tests in LN$_2$ and LAr. More investigations have yet to be performed to verify the feasibility of deploying such a technique for a physical experiment. Collaborative research among institutions with complementary expertise and resources would largely accelerate the progress in this interesting and important direction.

\begin{acknowledgement}
  The authors would like to thank Mark Amman for his instruction on fabricating planar detectors, and the Nuclear Science Division at Lawrence Berkeley National Laboratory for providing the vacuum cryostat.
\end{acknowledgement}

\bibliography{refs.bib}
\bibliographystyle{spphys}

\end{document}